\documentstyle[aps,prl,preprint]{revtex}
\begin{document}

\newcommand{\be}{\begin{equation}}
\newcommand{\ee}{\end{equation}}
\newcommand{\eps}{\varepsilon}
\newcommand{\z}{\zeta}
\newcommand{\cc}{\xi}

\draft

\title{An exactly solved model of biological evolution}
\author{Stefano Galluccio}
\address{Institut de Physique Th\'eorique, Universit\'e de Fribourg,
Perolles, CH-1700, Fribourg, Switzerland}
\maketitle

\begin{abstract}
We  reconsider the Eigen's quasi-species model for competing
self-reproductive macromolecules in populations  characterized by 
a single-peaked fitness landscape. 
The use of  ideas and tools borrowed from polymers theory and 
statistical mechanics, allows us to exactly solve the model for generic
DNA lengths $d$. The mathematical shape of the quasi-species 
confined around
the master sequence is perturbatively  found  in powers of
$1/d$  at large $d$.
We rigorously prove the existence of the error-threshold phenomenon
and study the quasi-species formation in the general context of 
critical phase transitions in physics. 
No sharp transitions exist at any finite $d$, and at $d\rightarrow
\infty$ the transition is of first order. The typical r.m.s. 
amplitude of a quasi-species around the master sequence
is found to diverge algebraically 
with  exponent $\nu_{\perp}=1$ at the transition to the delocalized 
phase in the limit $d\rightarrow \infty$.
\end{abstract}

\section{Introduction}

In recent years there has been an increasing interest of theoretical
physics
 in respect to  new  interesting  phenomena
 for  which the general approach 
of statistical mechanics has turned out to be extremely powerful.
Typical examples are represented by earthquake  modelization
\cite{ofc},
forest-fire propagation models \cite{ds}, financial systems 
and stock markets dynamics \cite{bou},
portfolio theory \cite{gz} and population dynamics \cite{abr}. 

In the large context of biological models of evolution, the 
so-called {\it quasi-species model}, as first introduced by
Manfred Eigen \cite{eigen}, has to be considered
 the paradigm of all  systems describing  
the dynamics of competing macromolecular organisms. 
It mostly relies on Darwinian's natural selection principle
as the best suited general theory to explain the evolution of species
and their competition for life.
In general  it  is believed that this  principle has not only
guided species to their present level of evolution, but also acted
at a molecular level in order to create the first living beings. 
The complexity of life  as it is,  still represents a hard challenge for
the scientists. 
The natural questions arising  in this context  are
usually: i) how  is  it possible that among the huge number
of possible (stable) molecular structures, natural selection has
chosen  the ones appropriate for the appearance of life on our planet?
ii) Why this final state is so stable and perfect despite the 
number  of  possible casual mutations that can occur during
 evolution?
 If we count the number of different alternative DNA sequences 
that one   obtains by modifying a chain of given length,
 we would discover  that it is so huge that we are 
necessarily forced to admit  that the majority of the chemical
 combinations has  never being tested by natural evolution.

In this article we reexamine the Eigen's model in the simplest 
formulation, with a sharply peaked fitness landscape
on a  lattice.
By means of a mapping to an equilibrium problem, we solve the
model under  very general assumptions, and we discuss 
the consequences of our results in  more realistic 
situations.

The remainder of this paper is organized as follows. In Sec. II  and III 
we  give a short survey of  the quasi-species model as first formulated
 by Eigen and coworkers. 
Sections IV and V are devoted to the introduction of   our 
simplified lattice model. More specifically, we will
show how  the  Eigen's equations  can be mapped 
into  the statistical mechanics of directed
polymers in a random medium.
In Sec. VI we introduce the effective transfer matrix associated to
the system. It  will be used  to get  some preliminary  
analytical results.
Sections VII and VIII contain the basic ingredients towards a full 
solution of the problem: the dual space  method and the 
characterization of the error-threshold phenomenon as a 
thermodynamic phase transition.
Finally, in Sec. IX, we get the complete solution of the model
after summation of the partition function associated. 
The critical properties at the error threshold are  calculated.
A survey of the main results and a comparison
with previous approaches are  finally summarized in Sec. X.

\section{The quasi-species model.}

In order to look for a mathematical transcription of  Darwinian
theory we must first resume the basic statements of
natural selection: 
i) Life came about through evolution;  ii)
 Evolution is the result of mutations for thermodynamic 
systems out of equilibrium; iii)
 Mutations are due to noncorrect reproductions or errors during
the process.

The selective principle, sometimes called
``survival of the fitness'',  is actually  opposed
to coexistence among individuals. Even though 
 the fitness landscape had
strong fluctuations, evolutions would not   proceed very far
if it were based on correlations among species instead of competition.
Without a true competition for life, evolution would have needed
a much larger time (perhaps larger than the life of the Universe!)
to explore the advantageous mutations 
among the huge number of different choices in the fitness landscape.

Darwinian principle   is nothing but a sort of deterministic process
of  selection of the fittest individuals among all others with the 
implicit assumption that an advantageous mutant  can occur 
{\it by chance} during 
reproduction.  
This is, however, not the whole story. As demonstrated by Eigen
and coworkers in their famous work on species evolution
\cite{eigen}, some
guidance principle towards the advantageous mutants does exist,
as fitter species have more chance to appear that disadvantageous
ones.
In Darwinian models evolution is guided towards the peaks of
the fitness landscape, that is, even though no correlation
exists between a mutation and the fitness of the resulting mutant,
there is a tendency provided by the fact that the distribution
of mutants is fitness dependent and (statistically) not all 
mutations have  same probability to occur.

We say that two mutants belong to the same species if  at each
position of the DNA chain the found symbol is the prevailing one.
In a virus chain, $10^4$ single position errors can be present. If their
probability is uniform, the wild-type sequence would be, in average,
exact with probability of about 0.9999. 
In other words, at each site of a DNA chain one could find the
same nucleotide by  averaging among all the individuals of a given
 species
with an error of the order $\sim 10^{-5}$, even though each
mutant can have its own sequence which is different from those
one of the  others!
The target of the selection is therefore, not a single  individual,
but a set of mutants whose DNA chain is close, in the statistical sense
above defined, to that of a wild-type one.

Let us now introduce the Eigen's model. 
Imagine that each individual is defined by a DNA chain 
and consider all individuals  having a chain of same length $d$.
For each site of the chain in the primary structure, we can have
$k$ different nucleotides which appear in a random manner.
In a DNA  or RNA structure they can be of 4 different types
(G, A, C, U).  Alternatively, to simplify the problem, we can decide
to distinguish only among purines (R) and pyrimidines
(Y); in the latter case we assume $k=2$.
The total number of  possible  sequences of purines and pyrimidines
is given by $M=2^d$, and results in a extremely big number of
choices.  A single ribosomal RNA (for which $d=120$) is one
on $10^{72}$ possibilities, and a viral genome (typically $d\sim 5000$)
is one among the $M\sim 10^{3000}$ alternative sequences.
For more complex  species this number increases even wildly and
one can appreciate the order of magnitude of  the  typical numbers
involved in the system.
In the 
statistical mechanics language,  these systems must be 
represented in a discrete  phase space with volume of the 
order of $10^{10^{4,5}}$.

In order  to mathematically define affinity among individuals and
species, we need a quantitative measure suitable for mathematical
description.  This can be achieved by introducing the 
{\it Hamming distance} $D_H$. It is defined as follows: given two
individuals $I_i$ and $I_j$ each having its own sequence of length
$d$, their Hamming distance is given by the number of different
positions which are occupied by different basis (G,A,C or U).
Two individuals having a smaller $D_H$ than another couple,
are also more biologically affine.   

A correct classification of mutants according to their Hamming distance
requires  a space of dimension $d$ in  which each dimension 
consists of $k$ sites. Mathematically, the configuration space $\Omega$
is a $d$-dimensional hypercubic lattice in which each side contains
$k$ identical sites.  In  the simplest case of  only two kinds of basis,
($k=2$)  each site has a 1-to-1 correspondence with binary sequences.
Therefore each point of $\Omega$ represents a given wild-type
and its neighbors the mutants  with closest biological affinity.
We assume  to assign to each site ${\bf x} \in \Omega$
a variable, or discrete field ${\cal Z}({\bf x})$,  giving
the relative concentration of wild-types of kind ${\bf x}$ in the 
total population.

The topological structure of $\Omega$ has interesting  properties.
By increasing the dimension $d$,  the number of different
ways by which two points in $\Omega$ at distance $L$ 
can be connected increases much faster (as $L!$) than the
number of points having that distance, whose number goes
as $2^L$. 
This has the effect  that, if $d$ is large,  an enormous amount
of  sites are confined among them with a relative small Hamming
distance.  Biologically this means that in the ``genome space''
$\Omega$,  even small mutations (e.g. one-basis error reproductions)
can yield, after short time, to explore a big region  in the whole
accessible space, of  total dimension $2^d$.

Moreover, as the number of different paths is of the order $L!$,
a given species can  easily  transform into another one
by avoiding unfavorable  ways (e.g. disadvantageous species).

Finally, in the very  general situation, we must  assign
to each site in $\Omega$ a  variable identifying the fitness
of that given sequence.  This quantity must be a frozen variable,
that is its value must be conserved during evolution, as it 
schematically represents the quality of reproduction of that 
particular DNA sequence. From the mathematical point of view, 
the fitness landscape is represented by a rough  function
and defined by 
quenched random variables. This fact  has the effect of rendering
the solution of  the model a very hard task, like
in the spin glass problem \cite{franz}.

In his simpler formulation the sequences are self-reproductive, i.e.
individuals reproduce themselves asexually, and mutants appear
through mutations of  their respective parents.
We then introduce  a random variable with uniform distribution
in $[0,1]$,  the {\it copying fidelity} $q_i$.
From  experimental observations, the typical values
of $q_i$ are very close to 1, that is the probability that
a given reproduction process creates a mutant different from
the original  parent is very small. From simple combinatorics
 we get that
the probability that  successive consecutive mutations 
bring a  species $I_i$ to a different  $I_j$ (whose reciprocal Hamming
distance is  $D$) will be
\be
Q_D=q^d\left(\frac{ 1/q-1}{k-1}
\right)^D.
\ee  
The mutation matrix ${\bf Q}=(Q_{i,i} |\; i,j=1,2,\cdots,k^d)$,
has elements $Q_{i,j}$ giving the probability of mutation
between $I_i$ and $I_j$. The reader should note that this approach
allow for different single-base mutations per time step. 

Let us introduce the dynamics by considering the following
hypothesis: i) Sequences reproduce themselves in a constant
 fashion and,
if any individual is present with concentration $n_i(t)$,
the rate of change of the population is given by $\dot{n}_i(t)$;
ii) Sequences generate by asexual reproduction with erroneous
replication and the rate depends linearly on the relative concentration.

The most general natural 
evolution equation for the  concentrations $n_i(t)$ of the 
species $I_i$,  will  then be given by \cite{eigen}

\be 
\dot{n}_i(t)=\sum_{j=1}^{k^d} W_{ij}\, n_j(t)\qquad
{\rm with}\quad W_{ij}=Q_{ij}A_j-\delta_{ij}D_i.
\ee 
In the above formula  we have introduced the rate matrix 
${\bf W}$ which contains both diagonal and off-diagonal terms.
$A_i$ are autocatalytic amplification factors, that is the relative
rates of replication of the species $I_i$. They  equally describe the
{\it fitness} of the respective individuals, as favorable species
generate a higher number of offsprings.
The diagonal terms $W_{ii},\, (i=1,\cdots,k^d)$,
 correspond to reproduction processes
involving perfect replication of sequences, while off-diagonal 
terms to mutations  of the original ancestor. 
In order to maintain the total population constant, one has to
take into account external constraints causing the spontaneous death
of individuals. This can be simply achieved by summing to the 
diagonal terms the {\it decay rate} $D_i$ of the species $i$
(counting the number of deaths per unit time). Its inverse is the
average lifetime.

It is worth to point out that both $A_i$ and $D_i$ are (in general) 
quenched
variables in the equations.  Each species $i$ is supposed 
having a given fitness and decay rate, fixed by external condition
and by genetic informations.  These parameters 
must be considered as ``frozen'' during evolution.

\section{Guided evolution and error-catastrophe.}

 Eigen and coworkers were able to show that 
 this simplified level of  description is indeed well defined
if the concentrations  $n_i(t)$ are not too high, and  the 
replication  rates $dn_i(t)/dt$ linearly depend on the 
concentrations themselves. 
At higher densities, the solution saturates and the creation of new
templates happens in more complex forms (for a review
see \cite{eigen}).
Even taking into account  these effects, the proposed model can
be shown to stay valid at a qualitative level of description,
as the system still have rates that linearly depend  (in average)
on the concentrations.
There are, however, situations in which a  linear model  cannot
describe the actual reproduction mechanisms.
A virus can, for instance, reproduce in the early stages of an 
incoming infection at much higher rates than those described
by Eigen's linear model. 

We  are now ready to a  deeper  investigation
of the  Eigen's model. 
To this aim it is advantageous to introduce a  rescaled quantity
\be
x_i(t)=\frac{n_i(t)}{\sum_{j=1}^{k^d}n_j(t)},
\ee
which represents the fractional population variable.
In its complete form we should add to eq. (2) a term which 
takes into account for changes in the population caused by transport
effects. To this aim one usually  introduces
 a general ``flux'' term $\phi(t)$ to account for any external
restriction on the total number of individuals.
We  can thus write the kinetic equations as
\be
\dot{x}_i(t)=\sum_{j\ne i} W_{ij}x_j(t)-\phi(t)x_i(t).
\ee
If one neglects  $\phi(t)$, 
the above equation simplifies into 
 a high-dimensional linear differential system 
whose matrix ${\bf W}$ is supposed to be diagonalizable.
(This hypothesis is believed to be satisfied in general situations).
As, moreover, ${\bf W}$ is definite positive, Frobenius theorem applies,
that is the maximum (or {\it dominant}) eigenvalue 
$\lambda_0$ is positive and nondegenerate,
and has a corresponding positive eigenvector.
It gives the net production rate of sequences in the stationary state,
and the corresponding (positive) eigenvector 
$(x_1,x_2,\cdots,x_N)$ is associated
to the relative concentrations of individuals in the total of
the population.
Formally, the  full stationary solution is a 
superposition of  uncoupled modes 
 and  in the limit of large times
the evolution is associated to the eigenvector corresponding 
to  $\lambda_0$.

It can be  shown that the average eigenvalue $\overline
{\lambda (t)}$ acts as a threshold: modes corresponding to
$\lambda_i > \overline {\lambda (t)}$ grow indefinitely during
evolution, while modes with $\lambda_i < \overline {\lambda (t)}$
die out. 
Each normal mode corresponds,  in the original variables $x_i(t)$, 
to a set of sequences (or a ``clan'') with high biological affinity.
A clan is uniquely defined by an eigenvector  and
its  associated eigenvalue. It competes for selection with
all other clans and  the target of evolution is the group corresponding
to $\lambda_0$. 
If viewed in the original space, a clan is represented by a set of
sequences distributed around the  one corresponding to the
largest diagonal term $W_{ii}$ which will be called 
{\it master sequence} (MS). The  mutants of the MS are grouped around
it  in such a way that only their averaged sequence equals that 
of the MS itself, which will be thought of as the much abundant
individual  in the set  (though variances can be  very large
around the MS). This set is called {\it quasi-species}.
  
The picture  that emerges from the above considerations is that 
of a huge number of individuals transforming one into the other during
evolution.  After some time all individuals will be found to be close
to a limited number of MSs,  as  less favorable species 
have already died out. 
The characteristic time necessary to reach a unique MS starting
from a flat distribution in the space of sequences is not infinite,
despite the number of sites in the system. 
This is due, as  previously pointed out,
to the topological  structure  of  $\Omega$, 
in which points very far apart can be reached in few steps and are
linked each other by a tight network of different paths.
As a consequence, a given sequence will almost certainly find
a more favorable region in the rugged  landscape
by  performing  a walk in $\Omega$ that avoids passing through
high potential barriers where it would stay blocked for long time. 

This principle of {\it guided evolution}  depends on the
off-diagonal terms of the matrix ${\bf W}$.  If they are zero,
no mutations occur and the global population is stationary.
If they are too big respect to the diagonal terms $W_{AA}$,
the ``diffusion'' in $\Omega$ is overenhanced and the stationary
state is dominated by a random creation and annihilation 
of all sequences. In this  situation the typical  spatial amplitude 
of a quasi-species becomes  of the same order of $d$ 
and no MS can be uniquely defined. We would reach the same
final state if  the fitness landscape would be  flat, i.e. $A_i=const.$
$\forall i$.
As a consequence,  we deduce that it may exist a critical value of
the error rate $q_c$ such that if $q <q_c$  the class of sequences 
classified as fittest becomes so large that it cannot be sampled by
any biological population.

This phenomenon was indeed  shown to exist for a large variety 
of fitness landscapes \cite{eigen} and it is now well accepted as 
an intrinsic feature of the quasi-species model. 
A rough estimate of 
$q_c$  (usually called  {\it error threshold}) can be achieved 
by noting that in order 
for a given sequence $I_i$ to be competitive with
other mutants,  its  exact replication
rate $W_{ii}$ must be larger than the average production rate
 of the mutants $\overline{E}_{j \ne i}$. 
On this basis it is possible to show \cite{eigen}, that  the 
conditions reads
\be
W_{ii} >\overline{E}_{j\ne i}=\frac{\sum_{j \ne i}E_j\overline{x}_j}
{\sum_{j \ne i} \overline{x}_j},
\ee
where $\overline{x}_j$ are the stationary relative concentrations
of the mutants.
Since, by definition, $W_{ii}=A_i Q_0-D_i$ and $Q_0=q^d$ 
is the probability of exact replication, we find that the 
critical threshold  reads
\be
Q_0 > \frac{\overline{E}_{j\ne i} +D_i}{A_i} =\frac{1}{\sigma}.
\ee 
Hence it follows that, in order to have localization around the MS,
 the length of the sequences must not exceed the critical
value
\be
d_{{\rm max}}=-\frac{\ln \sigma}{\ln q}\sim \frac{\ln \sigma}{1-q}\,
\quad {\rm for}\quad 1-q \ll 1.
\ee
Once that  both $q$ and $\sigma$ are fixed,  we then have a severe
restriction on the maximum possible length which allows 
selection to find the optimal MS. 
The above condition can be equally 
rewritten in  terms of the autocathalitic
rate as 
\be
A_i > \left( \overline{E}_{j\ne i}+D_i\right)
\left(\frac{1}{q}\right)^d \sim e^{a d}. 
\ee
The last inequality can be expressed by saying that in order
to maintain a given quasi-species  stable around a MS  one needs
the corresponding selective advantage (or fitness) to exceed
a given threshold. 
What is surprising is the functional dependence of this threshold
on the length of the sequences: since typically $d$ if the order
of $10^{3, 4}$,  the minimum $A_i$ requested is enormous!
Fortunately, this is not devastating, because of the presence of
the factor $1-q \ll 1$ at the denominator in (7).
   
\section{Towards a  solvable model of evolution.}

A full complete solution of the Eigen's model is not achievable
by analytical methods, and despite  past extensive work 
\cite{af96,let87,t92},  no exact solutions are still available
in the literature. 
An important result in this context was achieved by Leuth\"ausser
\cite{let87}, who first showed the link between the quasi-species model
and the statistical mechanics of lattice surface systems.

Our goal is to introduce  a simplified version of  the Eigen's equations
which, although  being well suitable for analytical approach,
still retains the basic fundamental features of the general system.
In particular we will consider a model  in discretized
time, as in \cite{let87} and, after having exactly solved the problem 
for generic sequence lengths $d$,  we will  prove that the transition
 from a localized quasi-species  to a random distribution of individuals
is equivalent to a first order phase transition.
The mapping is based on the observation that the  system
admits a simple representation in terms of equilibrium
statistical physics. 
Similar ideas were already introduced by \cite{let87},  in which 
the  main idea was to map of the ODE (4) into 
a multidimensional Ising-like spin system at equilibrium.
However, due to the complex form of the ``effective'' Hamiltonian
resulting from the mapping, which contains
a complicated interaction term depending on the selective
advantages $A_i$,  this approach allowed only 
for  numerical  solutions. Tarazona \cite{t92} performed, on this
basis, a series of interesting computations with
different fitness landscapes and found a rich resulting scenario.

Our idea is  to introduce  a  different  mapping of the Eigen's equations 
to an equilibrium  statistical system, which, 
in our opinion, is simpler and more natural 
than the  one used in \cite{let87}.
By means  of this new mapping, in fact, we can directly 
relate  eq.(4) to a well-known problem in statistical
mechanics, that is,  directed polymers in random media (DPRM)
\cite{hhz}.
Due to the large amount of work done in this domain in the past years
\cite{fln}, a mapping to DPRM is important for many  reasons.
First of all, the physics of DPRMs  has applications in a large variety
of physical phenomena, and  it would be at least interesting
to compare all these systems with the evolutionary dynamics
proposed by Eigen.
On the other hand, due to the large amount of analytical 
and numerical work done in the  directed polymers context, we 
have a solid background which can be used to understand,
on   a more rigorous basis, 
 the physics  behind the quasi-species model.

In particular, in this paper, we will concentrate 
on the   characterization of the 
error-threshold phenomenon as a phase transition, and 
the calculation of the critical exponents involved (in the simplest
case we have considered).
Anticipating future conclusions, the error-threshold 
transition turns out to be equivalent 
to a depinning phase  transition of a directed polymer 
by a bulk potential \cite{fln}.
For sake of completeness, in the last section, we will discuss our 
results in respect to those obtained by previous approaches.

In order to introduce our model, we first formulate some
general  hypothesis.
\begin{enumerate}
\item We consider sequences defined by two-state basis (e.g. Y and R);
that is we take $k=2$. Each sequence of length $d$
 is made of a combination of ``0'' and ``1'' bits and $\Omega$
is the unitary hypercubic lattice $\{0,1\}^d$.
\item The fitness landscape is flat but one point (take the origin
${\bf 0}$)    having higher fitness. In other words we consider a
single-peaked distribution of selective advantages, by taking
$A_i=b$, if $ \Omega\ni{\bf x} \ne {\bf 0}$ and
$A_i=a >b$, if  $\Omega \ni {\bf x}={\bf 0}$.
\item
The decay rates are zero, i.e. $D_i=0$,
 $\forall i=1,2,\cdots,k^d$.  We have numerically verified that
this assumption does not affect our final conclusions.
\item We consider evolution in discretized time.  Eigen's model
is (formally) similar to a  system of coupled master equations in 
the variables
$x_i(t)$ if we interpret $x_i(t)$ as the ``probability to find a 
 localized quasi-species around the MS $I_i$ at time $t$''.
If we imagine to consider  the time as   a multiple of a small
interval (or {\it waiting time}) $\tau$, i.e. $t=N\tau$,  
we can write that
\begin{eqnarray}
\dot{x}_i(N\tau)&=&\lim_{N\rightarrow \infty}
\frac{x_i((N+1)\tau)-x_i(N\tau)}{\tau} \nonumber \\
&\stackrel{N \gg 1}{\sim}&\frac{1}{\tau}\left(
\tilde{T}_{ij}-\delta_{ij}\right) x_i(N\tau).  
\end{eqnarray} 
Usually $\tau$ is simply related to the inverse of the transition 
probability per unit time in the continuous  equation.
The above relation shows that, apart from the identity operator
$\delta_{ij}$, the dynamics on the discrete time can be described
by the repeated application of  a  $2^d\times 2^d$ 
{\it transfer matrix} $\tilde{T}_{ij}$ with $i,j=1,2,\cdots,2^d$. 
\item 
In general,  one should take into account multiple one-basis mutations
per time step $\tau$. This is contained in original Eigen's model 
as the rate matrix $W_{ij}$ has all non zero off-diagonal entries.
Nevertheless, 
 we will formulate the hypothesis that the transfer matrix
$\tilde{T}_{ij}$ can be reduced to another matrix $T_{ij}$
which allows only single-basis mutation per time step. 
The reason is that $T_{ij}$ has a much simpler structure than
$\tilde{T}_{ij}$, since almost all off-diagonal elements are zero.
We will prove below that  using the one-jump formulation
of the system does not modify the physical picture that
emerges from the model. In fact, allowing more than one mutation
per time step,  corresponds to take higher powers of $T_{ij}$,
as one can easily see.
All our results can be associated, however (see below),
to the behavior of the set of eigenvectors
 of the transfer matrix which does not depend
on the power of $T_{ij}$ we actually take into account.
\end{enumerate}     

We  finally  note that, without loss of generality, one  can take $b=1$,
apart from unimportant multiplicative factors.
 
\section{The model.}

Let  us consider a $d$-dimensional 
hypercubic  unitary lattice $\Omega=\{0,1\}^d$, 
representing  the  configuration  space.
For mathematical convenience, we will assume to have periodic
boundary conditions in all directions, even though this hypothesis
is not essential to the physics of the problem. 
Each side of $\Omega$  is made of only two points  representing
binary units. Each  point of  $\Omega$ has a one-to-one 
correspondence to a sequence $I_i  \quad ( i=1,|{\cal I}|)$
since  the cardinality of ${\cal I}$ is equal to the number
of  points of $\Omega$.
We formulate
 the implicit  hypothesis that all individuals of the population
have the same sequence length $d$.

On each site ${\bf x} \in \Omega$ we have a variable ${\cal Z}({\bf x})$
corresponding to the relative concentration of individuals of
wild-type $I_{{\bf x}}$.  Equivalently,  we can interpret
${\cal Z}({\bf x})$
 as the probability to find the sequence 
$I_{{\bf x}}$ in the total of the population. 
At each time step a fraction $t \in [0,1]$ of the population 
of the same wild-type  reproduces incorrectly  and their 
sequences change one basis among  $d$ and  transform itself 
into  a new set  of individuals  $I_{{\bf y}}$.
In our usual probabilistic interpretation, $t$ gives the
probability that  the MS $I_{{\bf x}}$ transforms into $I_{{\bf y}}$.

Since there are $d$ basis for each sequence, the probability that
a mutation takes place  is  $dt$ while $1-dt$ is the probability of exact
replication. In other words, $1-dt$ is the fraction of the 
population  ${\cal Z}({\bf x})$ which survive evolution. We need 
to consider pairs  ($d,t$) such that $dt <1$.
This is not a limitation of  our approach, in fact, even though 
$d$ is usually very large,
we  only study conditions in which the reproduction fidelity is
very high, i.e. $t \ll 1$.

All sequences have the same fitness $b=1$, apart from the origin
${\bf 0}=(0,0,\cdots,0)$ having  selective advantage $a>1$.

It is then simple to write down a recursive relation for the 
relative concentrations ${\cal Z}_N({\bf x})$ at time $N$  on the basis
of the above arguments:
\begin{eqnarray}
{\cal Z}_{N+1}({\bf x})&=&
\left(1+(a-1)\delta_{{\bf x}, \vec{0}}\right) \nonumber \\ 
&\times& \left(
\sum_{i=1}^{d}  t{\cal Z}_{N}\left({\bf x}+{\bf e}^{(i)}\right)+
(1-dt){\cal Z}_{N}({\bf x})\right),
\end{eqnarray}
where we have introduced the unitary vectors 
${\bf  e}^{(i)}$ as those having 
a``1" bit  as  $i$th element if ${\bf x}$ has a ``0'' in the same
position and viceversa.
The above  equation uniquely defines  the transfer matrix
$T_{ij}$ as ${\cal Z}_{N+1}({\bf x})={\bf T} {\cal Z}_N({\bf x})$.

The interpretation of the above relation is simple.  At time $N+1$,
the fraction of individuals with sequence $I_{{\bf x}}$ is equal
to $(1-dt)$ times the original concentration
${\cal Z}_N({\bf x})$ (this corresponds to the individuals who have not
experienced any mutation), plus the fraction of individuals 
with Hamming distance equal to 1 respect to ${\bf x}$ 
who, after reproduction,  have mutated to $I_{{\bf x}}$.
This fraction is  given by $t{\cal Z}_{N}\left({\bf x}+{\bf e}^{(i)}\right)$.
Moreover, we have chosen the origin as a favored species,
that is the population in ${\bf x}={\bf 0}$ is amplified by a factor
$a>1$ respect to all others. 
This hypothesis is nothing but a simple mathematical way to impose
that  there exist a {\it single } MS  $I_{{\bf 0}}$.
 
In this framework, the existence of a quasi-species characterized
by a unique MS corresponding to  $(0,0,\cdots,0)$
depends on its selective advantage respect to other sequences,
i.e. on the value of $a$. 
 We  thus expect 
to find  quasi-species formation  around $I_{{\bf 0}}$ if
$a$ is larger than a threshold $a_c$. 

Roughly speaking, this transition can be equally interpreted in a 
different context. 
Let us indeed  consider a directed elastic polymer (a line) wandering 
in $\Omega$, directed along the ``time''  axis $N$, and subjected
to an attractive potential located at the origin ${\bf 0}$.
If the potential is uniform in $N$, the  energy gain per time
step located at the wall is $-U$.   
If we introduce a vector ${\bf h}^{(i)}\in  \Omega$,  we can use it
to identify the position of the polymer at each time step $i$.  
The elasticity of the polymer, in a discrete geometry, is usually
described by restricting the one-step polymer 
fluctuations to be smaller than a fixed threshold.
In the literature this constraint is usually called RSOS condition,
and means that $\left|{\bf h}^{(i)}-{\bf h}^{(i-1)}
\right|$ can be 0 or 1 \cite{fln}. 

In a continuous formulation, the polymer statistics is associated
to a restricted (i.e. with fixed extremes) 
partition function (here ``s'' is the continuous
analogue of $N$)
\begin{eqnarray}
{\cal Z}({\bf h},s)&=&\int _{{\bf h}(0)={\bf 0}}^{{\bf h}(s)={\bf h}} 
{\cal D}\left[{\bf h}'\right]\;  \\ 
&\times &\exp\left\{ -\beta \int_0^s d\, s'
\left[ \frac{\nu}{2} \left(\partial_{s'} {\bf h}'(s')\right)^2
+V({\bf h }',s') \right]\right \}. \nonumber
\end{eqnarray}
In  general, $V({\bf h },s)$  is a random potential distributed according
to  
a given density (DPRM problem).
In the discrete formulation, we introduce
 a  Hamiltonian  with short-range uniform interaction
\begin{equation}
{\cal H}_N\left(\{{\bf  h}\}^{(i)}\right)=
\sum_{i=1}^{N} \left(J\left|{\bf h}^{(i)}-{\bf h}^{(i-1)}\right|-
U\, \delta_{{\bf h}^{(i)},{\bf 0}}  \right),
\end{equation}
as our potential is localized at the origin and is attractive, that is
$V({\bf h}^{(i)},i)=-U\delta_{{\bf h}^{(i)},{\bf 0}}$.
The continuous partition function then becomes a sum 
over all possible realizations of the  restricted polymer
between $0$ and $N$ \cite{hhz}
\begin{equation}
{\cal Z}_N({\bf x})=\sum_{\{{\bf h}\}} \exp\left\{ 
-{\cal H}_N\left(\{{\bf h}\}^{(i)}\right) /T\right\}.
\end{equation}

The above sum completely specifies the state of the polymer at a given 
temperature $T$, or equivalently, at a given potential 
strength $U$.
By general considerations,   we know that in the thermodynamic
 limit the polymer has a phase transition from a localized
into a delocalized state, depending on $T$, or, equivalently,
on $U$.
As we will discuss below, this transition is perfectly defined only 
at $d\rightarrow \infty$, as the cardinality 
of $\Omega$ is finite for every finite $d$ and thermodynamic
limit does not hold.

There exists an interesting  mapping between the 
Eigen's model 
 and the   statistical mechanics of a directed polymer. 
For instance, in our case,
a simple look at the partition function (13) shows that 
it is mathematically equivalent to  the 
species concentration ${\cal Z}_N({\bf x})$ 
which  identically satisfies the recursive relation  (10), once 
we have introduced the definitions
$a=\exp(U/T)$ and $t=\exp(-J/T)$.
That is why we implicitly used the same notation for the 
concentration of  individuals and the polymer partition function.

As an example, let us suppose that for a given
set $\{a,d,t\}$ the  polymer is in the localized  (delocalized) 
phase: this can be  equivalently expressed by saying 
 that  evolution brings species
preferentially to  (apart from) the master sequence $I_{{\bf 0}}$.  
Therefore the error threshold transition in the 
self-reproductive model is reduced
to the search of  the critical pinning $a$  necessary 
to localize the directed polymer for fixed  values of  $d$ and $t$.
The error catastrophe transition will then be perfectly 
understood in the general context of  thermodynamic
 phase transitions.
Even though we will concentrate our study on the simplest case 
of a single peaked fitness, it is worth mentioning that   the same
formalism applies in more realistic  situations, for which
we are forced to 
consider a quenched bulk potential, as  in  (11).
Generally speaking, 
studying of the dynamics of the quasi-species model,
turns out to be not a  simpler problem 
 than   DPRM.  

\section{The effective matrix.}

In order to calculate the partition sum (13), 
we must  first solve a $2^d \times 2^d$ eigenvalue
problem associated to  the transfer matrix ${\bf T}$, 
${\cal Z}_{N+1}({\bf x})={\bf T}{\cal Z}_N({\bf x})$.

As we are interested in the stationary state at $N\rightarrow \infty$,
we don't need  to find the whole spectrum of ${\bf T}$
but  its spectral radius (i.e. the maximum  eigenvalue) $\eps$
as a function of the  free parameters $\{a,d,t\}$, only.
$\eps$ the only significant contribution to 
the free energy density (per unit length) 
$f=\lim_{N\rightarrow \infty}-\log({\cal H}-TS)/\beta N$.
At large times $N$,  the action of ${\bf T}$ is dominated 
by the spectral radius $\eps$, that is 
${\cal Z}_{N+1}({\bf x})\sim \eps {\cal Z}_N({\bf x})$.

It is worth considering some simple mathematical preliminaries which
will be useful in the future. 
A straight investigation of  the transfer matrix shows that  it is 
definite positive and irreducible, and then,
as a consequence, Perron-Frobenius
theorem on finite matrices applies \cite{lt}:  the spectral radius 
is positive and non-degenerate, and corresponds to a positive, unique,
eigenvector. 

Due to the high dimensionality of the system (recall that 
typically $d \sim 10^{3,4}$ in a virus sequence), it is  not 
convenient to use this form of the matrix for numerical 
investigation.

To this aim,  we observe that the system has a symmetry respect
to any change of ``1'' and ``0'' bits in a  given sequence.
In other words, if two points ${\bf x}$ and ${\bf y}$ of $\Omega$
have the same Hamming distance from the MS 
$(0,0,\cdots,0)$ they are completely equivalent. The transfer
matrix is in fact completely invariant in this case under permutation
of the two points. 

Therefore the partition function must be invariant under
rotations in $\Omega$ and we can restrict ourselves to
study its radial  dependence only, i.e. ${\cal Z}({\bf x})=
{\cal Z}(|{\bf x}|)={\cal Z}(\nu)$
where we have defined $\nu=D_H({\bf x},{\bf 0})$.
It is a simple combinatorial result  that the number of points
of ${\Omega}$ with the same Hamming distance $\nu$ 
from the origin is given by $M=d!/[(d-\nu)!\nu !]$.

If we define a new vector as $P_N(\nu)=\sum_{|{\bf x}|=\nu}
{\cal Z}_N({\bf x})$ we can equally study our eigenvalue
problem in terms of a new transfer matrix ${\bf S}$  defined as
$P_{N+1}(\nu)={\bf S} P_N(\nu)$.
It can be found by observing that
\begin{eqnarray}
P_N(\nu)&=&\left(1+(a-1)\delta_{{\bf x}, \vec{0}}\right)  \\
&+&
\left(
\sum_{|{\bf x}|=\nu}\sum_{i=1}^{d}  t{\cal Z}_{N}\left({\bf x}+
{\bf e}^{(i)}\right) \right. \nonumber \\
&+& \left.
 (1-dt)\sum_{|{\bf x}|=\nu}{\cal Z}_{N}({\bf x})\right), \nonumber
\end{eqnarray}
where the last term in parenthesis is 
simply given  by $(1-dt)P_N(\nu)$.
By definition,  ${\cal Z}_{N}({\bf x}+{\bf e}^{(i)})$ is of the form
${\cal Z}_{N}({\bf x})$ with $|{\bf x}|=\nu+1$ or 
$|{\bf x}|=\nu-1$. Hence, after some algebra, we find that
\begin{eqnarray}
\sum_{|{\bf x }|=\nu}{\cal Z}_N({\bf x}+{\bf e}^{(i)})
&=&
(\nu+1)\sum_{|{\bf x }|=\nu+1}{\cal Z}_N({\bf x}+{\bf e}^{(i)})
\\
&+&
(d-\nu+1)\sum_{|{\bf x }|=\nu-1}{\cal Z}_N({\bf x}+{\bf e}^{(i)}).
\nonumber
\end{eqnarray}
By using this identity we can show that the recursion relation 
for $P_N(\nu)$ reads
\begin{eqnarray}
P_{N+1}(\nu)&=&\left(1+(a-1)\delta_{{\bf x}, \vec{0}}\right)
\left[ (1-dt)P_N(\nu)\right.  \nonumber \\
& + &\left. t(\nu+1)P_N(\nu+1)+ (d-\nu+1)P_N(\nu-1)\right] 
\nonumber \\
&  & {\rm with} \qquad P_N(\nu)=0\quad {\rm for}\quad \nu > d.
\end{eqnarray}
 
We can then study the system by means of an {\it effective} 
$(d+1) \times (d+1) $ matrix ${\bf S}$ defined as
\be
{\bf S}=
\left( \begin{array}{ccccc}
a(1-dt) & at &  &  0 &\\
dt & (1-dt) & 2t &   &\\
& \ddots & \ddots & \ddots & \\
&  & 2t & (1-dt) & dt \\
&0  & & t & (1-dt) 
\end{array} \right ).
\ee
It is easy to see that ${\bf S}$ and ${\bf T}$ are
completely equivalent  to our problem, since they have same
spectral radius, as it turns out from very general 
results in group theory  \cite{pen82}.
  In advantage, ${\bf S}$ is certainly more suitable for numerical
diagonalization respect to ${\bf T}$. 

What is more important, however, is that  we can use the
effective matrix to calculate some accurate upper and lower
bounds for    $\eps$.
This is a consequence of a theorem on positive, irreducible
matrices: it states that 
the spectral radius $\eps(A)$
of a positive  matrix $A =a_{ij}$ satisfies the inequalities
\begin{itemize}
\item $\min_i \sum_j a_{ij} \le \eps(A) \le \max_i\sum_j a_{ij}$,
\item $\min_j \sum_i a_{ij} \le \eps(A) \le \max_j\sum_i a_{ij}$.
\end{itemize}
In summary, we find  that
\be
\eps({\bf S}) \ge\left\{
\begin{array}{c c }
1 &  \qquad a \le (1-dt)^{-1} \\
a(1-dt) &\qquad a \ge (1-dt)^{-1}
\end{array}\right.
\ee
while the upper bound is estimated as
\be
\eps({\bf S}) \le\left\{
\begin{array}{c c }
a(1-dt)+dt & \qquad  a \le \frac{1-(d-2)t}{1-dt} \\
1+2t & \qquad a \in \left (\frac{1-(d-2)t}{1-dt},
 \frac{1+2t}{1-(d-1)t}\right] \\
a(1-dt+t) & \qquad a \in\left( \frac{1+2t}{1-(d-1)t},d\right] \\
a(1-dt)+dt & \qquad a > d
\end{array}\right. .
\ee
The result is shown in Fig.(1) where  the two curves 
corresponding to the upper $\eps_+({\bf S})$
and lower bound   $\eps_-({\bf S})$ are plotted (dashed lines).
From the above inequalities we immediately find some interesting
results concerning our system. In fact  we deduce that, $\forall d$
finite,  $\eps>1$ and that  in the limit
 $d\rightarrow \infty$ the spectral radius is bounded between 
two values, converging to 
\begin{eqnarray}
\eps({\bf S}) &\rightarrow &1^+ \qquad {\rm if} \quad a 
<\frac{1}{1-dt} \nonumber \\
 \eps({\bf S}) & \rightarrow  & a(1-dt) \qquad {\rm if} \quad 
a>\frac{1}{1-dt}.
\end{eqnarray}
This result  indicates that $a=a_c=1/(1-dt)$ is the critical 
value of the pinning we need to localize the polymer at the 
origin for any  fixed set of parameters $(T,J)$. It is intuitively
clear that, rigorously speaking, we cannot have a phase transition 
at finite $d$, since finite is  the cardinality of $\Omega$, too.
Only in the limit $d\rightarrow \infty$ the polymers has a  finite
probability do completely delocalize from the defect;  at any finite
dimension it can wander  up to a  distance of the order of $d$
 even at 
$N\rightarrow\infty$. 
Naively speaking, we can say that, at large (finite) dimensions, and if the 
pinning strength is not big enough, the polymer is ``rough'' in the 
sense that
it can visit {\it all} accessible configuration space up to 
the maximum size allowed for that fixed $d$.
On the other hand, in the ``pinned'' phase, the transversal 
localization length $\ell$    within which the polymer 
is confined to the origin is independent on the linear 
size $N$ and is always finite (even at $d\rightarrow \infty$).
The two different behaviors take place at a given
characteristic value $U_c$ (or equivalently
$a_c$) of the pinning potential.
Later on we will further discuss this problem and its implications
in the biological context. 

It is worth noting that,  from simple inspection of the effective matrix,
one can also get some information about the distribution (or concentration)
of individuals in the configuration space.  This can be easily achieved by
the knowledge of the eigenvector associated to the spectral radius 
$\eps({\bf S})$. 
We  consider the sum of its components $m_N=\sum_{\nu=0}^d
P_N(\nu)$, and from the above iterative relation for $P_N(\nu)$
 have:
\begin{eqnarray}
&m_{N+1}=&\sum_{\nu=0}^d\left[(a-1)\delta_{\nu,0}+1\right]
\left[(1-dt)P_N(\nu) \right.\nonumber \\
&+& \left.t(\nu+1)P_N(\nu+1)+t(d-\nu+1)P_N(\nu-1)
\right] \nonumber \\
&=& (a-1)\left[(1-dt)P_N(0)+tP_N(1)\right] \nonumber \\
&+&(1-dt)\sum_{\nu=0}^d
P_N(\nu)+t\sum_{\nu=0}^d(\nu+1)P_N(\nu+1) \nonumber \\
&-& t\sum_{\nu=0}^d (\nu-1)P_N(\nu-1) +dt\sum_{\nu=0}^d P_N(\nu-1)
 \\
&=& (a-1)\left[(1-dt)P_N(0)+tP_N(1)\right]+\sum_{\nu=0}^d
 P_N(\nu).  \nonumber 
\end{eqnarray}
Apart from a constant multiplicative (normalization) factor, we find,
in the thermodynamic limit $N\rightarrow \infty$, that
\be
m=\frac{\eps}{\eps-1}\frac{a-1}{a}.
\ee
It is easy to  prove that the inverse of $m$ gives (apart from
a constant factor) the fraction of the population at the origin
(that is with MS equal to $(0,0,\cdots,0)$). In fact a simple calculation
shows that $m^{-1}\propto P(0)/\sum_\nu P(\nu)$.

The dependence of $m$ on  the pinning strength $a$ is depicted in
Fig.(2) in a semilogarithmic scale. We see that  $m\sim 2^d$ for 
$a<a_c$, i.e. the fraction of individuals  with MS  equal to ${\bf 0}$
is $2^{-d}$. In other words,  the origin is not, in this  situation,
a privileged site, as all  individuals are equally likely to be found
 in $\Omega$. 
In the opposite situation, at $a>a_c$, we see that $m$ is 
approximatively
given by $1$.  This means  that almost all the
population share the same sequence, the quasi-species is well
defined and evolution has reached a stationary state  around the 
master sequence $(0,0,\cdots,0)$.
Remarkably, the transition appears again to occur at
 $a_c =(1-dt)^{-1}$.

\section{Dual space approach.} 
The direct investigation of the effective transfer matrix has given some 
first insight in the physics of the problem, in particular with respect to
the origin of the phase transition.
The simplicity of our model fortunately allows an exact solution
which is, however, non-trivial,  due to the high-dimensionality of the 
system.

To this aim we first need to simplify  the transfer matrix ${\bf T}$
by means of an appropriate transformation.
As the system is defined on $\Omega=\{0,1\}^d$,
we use a discretized
transformation to achieve the result.
We then  introduce the  following 
dual space representation of the partition
sum ${\cal Z}({\bf x})$:
\be
{\cal Z}_N({\bf x})=\sum_{{\bf k}=\{0,1\}^d}
(-1)^{{\bf x}\cdot {\bf k}} {\cal Z}_{N}({\bf k}),
\ee
and its inverse
\be
{\cal Z}_{N}({\bf k})=\frac{1}{2^d}\sum_{{\bf x}=\{0,1\}^d}
(-1)^{{\bf x}\cdot {\bf k}} {\cal Z}_{N}({\bf x}).
\ee
The dual space  is obviously  identical to $\Omega$
and the Kronecker delta is defined as $2^d\delta_{{\bf x},{\bf 0}}
=\sum_{{\bf k}\in \Omega}(-1)^{{\bf x}\cdot{\bf k}}$.

A rapid inspection shows that this representation  implicitly contains 
periodic boundary conditions in all directions. In the dual space 
the transfer matrix ${\bf T}$ 
reads
\begin{equation}
{\cal Z}_{N+1}({\bf k})=s({\bf k}){\cal Z}_{N}({\bf k})+
\frac{a-1}{2^d}\sum_{{\bf q}=\{0,1\}^d} 
s({\bf q}){\cal Z}_{N}({\bf q}),
\end{equation}
with $s({\bf q})=t\sum_{i=1}^d (-1)^{q_i}+1-dt$.

Our goal is then to solve  a $2^{d}$-dimensional eigenvalue problem
for  the  dual transfer matrix ${\bf  T}$ acting on the
 r.h.s. of  the last equation.
In the limit $N\rightarrow\infty$ the system reaches a stationary state
and in this regime ${\bf T}$ is dominated by its 
 spectral radius $\eps$. We can then write that in the 
thermodynamic limit ${\cal Z}_{N+1}({\bf k})=\eps {\cal Z}_N({\bf k})
=\eps{\cal Z}({\bf k})$, and
\be
{\cal Z}({\bf k})=\frac{a-1}{2^d}\frac{1}{(\eps-s({\bf k}))}
\sum_{{\bf q}}
{\cal Z}({\bf q})s({\bf q}).
\ee
Let us focus, for the moment, to the computation of  $\eps$,
and  define a new  constant
$Q=\sum_{{\bf k}\in \Omega}s({\bf k}){\cal Z}({\bf k})$. 
By multiplying both sides by $s({\bf k})$, and summing over ${\bf k}$,
we finally arrive at the equation  for $\eps$:
\begin{equation}
\frac{a-1}{2^d}\sum_{{\bf k}=\{0,1\}^d} \frac{s({\bf k})}
{\eps-s({\bf k})},
\end{equation}
or
\be
\frac{a}{a-1}=\frac{1}{2^d}\sum_{{\bf k}=\{0,1\}^d} \frac{\eps}
{\eps-s({\bf k})}.
\ee

It is clear that any attempt to  directly calculate the sum appearing 
in the above formula is a very hard task, and we are forced  to
rely on
different approaches. 
First of all we note that, since $s({\bf k})$ can take values
of  kind $1-2nt$ (with $n=1,2,\cdots,d)$
 in $d!/(n!(d-n)!)$ different ways, we can   recast
the sum as follows
\be
\frac{a}{a-1}=\frac{1}{2^d}\sum_{n=0}^d
\left(
\begin{array}{c} d \\ n
\end{array}\right)
\frac{\eps}{\eps-1+2nt}.
\ee 
Despite the simplification, the last expression is still too hard to
be exactly  solved,  nevertheless it can be used  to study 
 the structure of the eigenvalues of  the transfer matrix.
In fact, the r.h.s. of the above equation has $d+1$  singular  points
in $\eps=1-2nt$, the largest of which is located at $\eps=1$.
For each interval between any two singular points, (29) behaves
as a continuous function of $\eps$ and it is monotonic decreasing,
then invertible. 
The solutions to the above equation are given by the intersections
of this function with the horizontal line $a/(a-1)$. There are 
$d+1$ intersections,  each of them corresponding to one eigenvalue
of the transfer matrix. 
As the largest singular point is located at $\eps=1$, we have a unique
eigenvalue larger than 1, and it corresponds to the 
spectral radius of ${\bf T}$.  We then 
 concentrate, in what follows,  on the solution  of (27)
with the restriction $\eps>1$, disregarding
all other roots. 
It is worth noting that in one simple case the sum  can be
explicitly performed. In fact if we  take $\eps=1+2t$,  the sum 
reads:
\be
\sum_{n=0}^d\left(
\begin{array}{c} d \\ n
\end{array}\right)
\frac{1}{1+n}=\frac{2^{d+1}-1}{d+1},
\ee
and (29) can be solved for $a$ giving 
\be
a=1+\frac{2t(d+1)}{(1+2t)(2-2^{-d})-2t(d+1)}.
\ee
Above we have anticipated that no sharp phase transition can occur
at any finite $d$. In performing the limit $d\rightarrow \infty$ we
 must be sure that $t$ goes to 0 at least linearly in $1/ d$
in  order to preserve the probabilistic interpretation of the system
(recall that   $dt \in[0,1]$).    
If, for instance, we suppose to approach the critical state on the manifold 
$\eps=1+2t$, that is $\eps\rightarrow 1$ linearly in $t$, we 
have, from (31), that $a\rightarrow a_c=1/(1-dt)$. 
This again proves  that, at least on the above manifold, 
$(1-dt)^{-1}$ is the critical selective advantage separating the two
phases. 

\section{The exact solution.}

Let us consider the eigenvalue equation (28). The idea is to
introduce a new representation to simplify the formula.
The bill we have to pay in this operation is that the final result
will be  expressed in implicit integral form.

By using a Feynman-like representation, we have
\begin{eqnarray}
\sum_{{\bf k}=\{0,1\}^d} \frac{\eps}
{\eps-s({\bf k})}& =& \frac{1}{A}\sum_{{\bf k}\in \Omega}
\left(1-\frac{1}{A\eps}t\sum_{i=1}^d (-1)^{k_i}\right)^{-1} \nonumber \\
&=&\frac{1}{A}\int_0^\infty du\; F(u,t,\eps,A),
\end{eqnarray}
with $A=(\eps-1+dt)/\eps$, and
\be
F(u,t,\eps,A)=
\sum_{{\bf k}\in \Omega}
\exp\left [u
\left(\frac{1}{A\eps}t\sum_{i=1}^d (-1)^{k_i}-1\right)
 \right]
\ee
By noting that the sum in the exponent
easily factorizes, we get 
\be
\sum_{{\bf k}\in \Omega}
\exp \left( 
\frac{u}{A\eps}t\sum_{i=1}^d (-1)^{k_i}\right)=
\left[2\cosh\left(\frac{ut}{A\eps}\right)\right]^d,
\ee
and therefore, after a change of variable in the integral, the eigenvalue
equation  takes the form
\begin{equation}
\frac{a}{a-1}=\eps\int_0^\infty  e^{-(\eps-1+dt)u}
\left( \cosh (ut)\right)^d\; du.
\end{equation}
A few remarks are important at this point on the meaning and
validity of the  above expression. It represents, for each fixed 
set of parameters $(T,J,d)$,  an integral implicit relation  between 
$\eps$ and $a$.  Nevertheless, it is not equivalent to the original 
series solution (28) of the spectrum of ${\bf T}$ since
in the above procedure we have implicitly assumed that
the integral representation were well mathematically defined.
In order to do this,  we must require that the integral (32)
 converges. This is indeed the case if and only if  
$\sum_{i=1}^dt(-1)^{k_i} <  A\eps$, or equivalently, if $\eps>1$.
If $\eps \le 1$ the integral diverges and no real solutions to the
above equation can be found. 
As a consequence, we can use eq. (35) to  calculate 
the spectrum of ${\bf T}$  corresponding to eigenvalues $>1$.
From the previous argument, we know that there exists a unique
eigenvalue larger than 1, and it corresponds to the spectral radius.
In conclusion, the unique  real solution in $\eps$ of (35) is
 the spectral  radius of the transfer matrix.
Moreover, since the integral  diverges  at  $\eps=1$,  
when  the  attractive potential at the origin
is omitted (i.e. $a=1$),  the maximum eigenvalue
  must be unitary, too.
Then the free energy density $f$ vanishes and we attain  a delocalized 
phase, as  expected.

The implicit integral can be  expressed in terms of known
mathematical functions.  After successive 
integrations by parts we have that (here $\delta=(\eps-1)/t$)
\begin{eqnarray}
\frac{1}{\eps}\frac{a}{a-1}&=&
\frac{1}{t\delta}\left(1-\frac{d}{\delta+2}\left(
1-\frac{d-1}{\delta+4}\left(1- \right.\right.\right.\nonumber \\
&\cdots& \left.\left.\left.
-\frac{2}{\delta+2d-2}\left(
1-\frac{1}{\delta+2d}\right)\cdots\right)\right)\right),
\end{eqnarray}
and recalling the definition of the hypergeometric series 
of negative argument \cite{as}
\be
F(-m,b;c,z)=\sum_{n=0}^m\frac{(-m)_n(b)_n}{(c)_n}\frac{z^n}{n!},
\ee
with $(a)_n=a(a+1)\cdots (a+n-1)$, we finally arrive at the result
that
\be
\frac{a}{a-1}\frac{\eps-1}{\eps}= F\left(-d,1;\frac{\eps-1}{2t}+1,\frac{1}{2}
\right).
\ee
We immediately deduce that (recall definition (22)) 
$m^{-1}=F\left(-d,1;(\eps-1)/2t+1,1/2
\right)$.

Let us define $I(d; \eps, t)$ the integral in (35). 
 $I(d; \eps, t)$ is a monotonic decreasing function
of $d$.  This result can be easily  proved by using the integral
representation of the hypergeometric series.
Physically we are interested  at the behavior of the system at large
dimensions, and in this regime we can use a Laplace saddle-point
 approximation
of  the integral solution. 
A detailed analysis of the asymptotic development 
of  $I(d;\eps, t)$ at large $d$  
needs  however particular attention, since we should 
properly  take into account the condition $dt\le 1$.
This means that  
both the limits $d\rightarrow\infty$ and $t\rightarrow 0$
must be performed {\it simultaneously} 
in such a way that $\alpha=dt$ be constant. We are implicitly assuming 
that $t$ goes to 0 linearly in $1/d$, but we would  
obtain
the same   final result
 if $dt\rightarrow 0 $ for $d\rightarrow \infty$.

Let $\alpha$ be equal to $dt$, a quantity which must be kept
finite during the calculation. 
We see that  $I(d;\eps, t)$ can be written as $\int_0^\infty
du \; g(u)\; \exp[d\, f(u)]$, with 
\begin{eqnarray}
f(u)&=& \ln \cosh(u)-\frac{\eps-1+\alpha}{\alpha}u, \nonumber \\
g(u)&=&\frac{d}{\alpha}.
\end{eqnarray}

Since for $\eps>1$ the maximum of $g(u)$ is located at the 
extreme of integration $u=0$, 
the  integral can be well approximated, at large $d$, 
by expanding in a McLaurin
series the integrand.  At first order in $1/d$ it reads
\be
I\sim \int_0^\infty du\;  g(0) \exp\left[d\left(f(0)+f'(0)u\right)\right]
=\frac{1}{\eps-1+\alpha}.
\ee
More precisely, if we take into account higher powers and the 
relative error, after  some more
algebra we arrive at the approximate result:

\begin{eqnarray}
\frac{1}{\eps}\frac{a}{a-1}&=&\frac{1}{\eps-1+dt}+\frac{(dt)^2}{d(\eps-1+dt)^3}
+\frac{3(dt)^4}{d^2(\eps-1+dt)^5} \nonumber \\
&+&\frac{1}{d^3}\left[\frac{15(dt)^6}{(\eps-1+dt)^7}-\frac{2(dt)^4}
{(\eps-1+dt)^5}\right] \nonumber \\
&+&O\left(\frac{1}{d^4}\right).
\end{eqnarray}
 
If we are interested in the  unique real solution of the above
algebraic  equation, (41) 
can  be inverted  
for the maximum $\eps$. The final solution, up to order $O(1/d^3)$
reads, for $a\in [1,\infty)$
\begin{eqnarray}
\eps&=&\max\left\{1,a(1-dt)+\frac{1}{d}\frac{a(dt)^2}{(a-1)(1-dt)}
\right.\nonumber \\
&+&
\left.\frac{1}{d^2}\frac{a(2-a)(dt)^4}{(a-1)^3(1-(dt)^3)}+O\left(\frac
{1}{d^3}\right) \right\},
\end{eqnarray}
since we know from the above arguments  (from
the effective matrix), that the spectral radius
cannot be less than 1 if $a>1$.
This result can be finally compared
with the exact calculation performed by numerically finding the 
spectral radius of ${\bf T}$ for a given  set of 
parameters $\{d, t, a\}$, and the two curves are plotted in Fig.(1).

In the limit $d \rightarrow \infty$  we have
\be
\eps^{(\infty)}=\max\{1, a(1-dt)\},
\ee
a result which coincides with that obtained from the analysis
we performed on the effective matrix ${\bf S}$.
Hence the critical selective advantage for the MS $(0,0,\cdots,0)$
to create a stable quasi-species around it, is $a_c=(1-dt)^{-1}$.
In other words, as we will clarify below, $a_c$ defines the error
threshold for  quasi-species formation.
Alternatively, one can arrive at the same result on the basis of
the convexity property of $I$ as a  function of $d$, 
as it  was showed in \cite{ggz95}.
Fig.(3) shows the critical dimension $d_c$ as a function
of  the pinning $a$ for two values of $t$.  The coincidence between
 (43) and the  numerical result is remarkable.
 
\section{The stationary ``ground state'' eigenvector.}

In order to have a full solution of our system, we still need to calculate
the partition sum (13), or more precisely, 
the eigenvector corresponding to
the maximum eigenvalue we have studied in the previous paragraph.  
Therefore,  let us go back to the recursion relation (25) in the 
dual space.
Disregarding, for the moment, the normalization condition,
we have ${\cal Z}({\bf k})=Q(a-1)2^{-d}/(\eps-s({\bf k}))$.
In the direct space, it reads
\be
{\cal Z}({\bf x})=Q\frac{a-1}{2^d}\sum_{{\bf k}\in \Omega}
(-1)^{{\bf x}\cdot{\bf k}}\frac{1}{\eps-s({\bf k})}.
\ee 
The summation of the series appearing in the above formula can be
done following  the same  general procedure  as before, that is,
 at $\eps>1$ we have
\begin{eqnarray}
\sum_{{\bf k}=\{0,1\}^d} \frac{(-1)^{{\bf x}\cdot{\bf k}}}
{\eps-s({\bf k})}& =& \sum_{{\bf k}\in \Omega}
\left(1-\frac{t}{B}\sum_{i=1}^d (-1)^{k_i}\right)^{-1} \nonumber \\
&=&\frac{1}{B}\int_0^\infty du\; G(u,x,t,B),
\end{eqnarray}
with $B=\eps-1+dt$ and
\be
G(u,x,t,B)=
\sum_{{\bf k}\in \Omega}
(-1)^{{\bf x}\cdot{\bf k}}\exp\left [u
\left(\frac{t}{B}\sum_{i=1}^d (-1)^{k_i}-1\right)
 \right].
\ee 
Respect to the above case,  we now have an
additional term in the sum over ${\bf k}\in \Omega$.
After  factorization, we find that
\begin{eqnarray}
&&\sum_{{\bf k}\in \Omega}(-1)^{{\bf x}\cdot{\bf k}}\exp\left(
\frac{ut}{B}\sum_{i=1}^d (-1)^{k_i}\right) \nonumber \\
&=&
\prod_{i=1}^d\sum_{k=0,1}(-1)^{kx_i} \exp\left((-1)^k \frac
{ut}{B}\right).
\end{eqnarray}
In this form the formula is  still too hard to allow a simple summation,
but a rapid inspection shows how to simplify the problem by
taking into account the symmetries of the system.
In fact we know that the partition sum must be the same for any 
two points with equal Hamming distance from the origin. Therefore
we can concentrate to study only the ``radial'' function $P(\nu)$
where $\nu$ is the Hamming distance from $(0,0,\cdots,0)$.
In practice this observation allows us to neglect   the order
in which bits ``1'' and ``0'' appear in (47).
What is   physically important is only the number of bits of each
kind which are contained in a given sequence of total length $d$.
If there are $\nu$ bits of kind ``1'', that is if the Hamming distance
of the respective sequence is $\nu$, in the product  at the r.h.s
of (47) will be present  $\nu$ factors of kind
$\exp(ut/B)-\exp(-ut/B)=2\sinh(ut/B)$ and 
$d-\nu$ of kind $\exp(ut/B)+
\exp(-ut/B)=2\cosh(ut/B)$. Finally, as the number of ways we can arrange 
$\nu$ bits ``1'' in the total of $d$ bits in $d!/[(d-\nu)!\nu!]$,
 we can write that:
\begin{eqnarray}
P(\nu)&=&Q(a-1)\left(
\begin{array}{c} d \\ \nu 
\end{array}\right)\int_0^\infty du\;
e^{-(\eps-1+dt)u} \nonumber \\
&\times & \left(\sinh(ut)\right)^\nu
\left(\cosh(ut)\right)^{d-\nu}.
\end{eqnarray}
The constant $Q$ can be fixed by normalization, that is, if we impose 
that ${\cal Z}({\bf x})$ be summable, we must require that
$\sum_{\nu=0}^d P(\nu)=1$. 
The last calculation is easy to be performed, in fact 
$\sum_{\nu=0}^d
\left(\begin{array}{c}
d \\ \nu
\end{array}\right)
 \left(\sinh(ut)\right)^\nu
\left(\cosh(ut)\right)^{d-\nu}=\exp(udt)$ and thus, after
integration, we get  the
result that $\sum_{\nu=0}^d P(\nu)=Q(a-1)/(\eps-1)$. The
normalized solution reads
\begin{eqnarray}
P(\nu)&=& (\eps-1)\left(
\begin{array}{c} d \\ \nu 
\end{array}\right)\int_0^\infty du\;
e^{-(\eps-1+dt)u} \nonumber \\
&\times &\left(\cosh(ut)\right)^d
\left(\tanh(ut)\right)^\nu.
\end{eqnarray}
At generic $d$ it is not possible to perform the above integral,
which is convergent $\forall \eps>1$, but we can   restrict
to study  the form  of the solution at large dimensions.

Since one may  equally characterize the depinning phase transition in
 terms of $U$ or  $a$, we can study   its order by considering
the discontinuities of the partition sum in $a$. 
At $d\rightarrow\infty$ the maximum eigenvalue is defined by
eq. (43). By inserting this expression into $P(\nu)$ we simply
find that the partition sum is a ${\cal C}^0$ function in $a$, that is
the phase transition is of first order.
This is also clear if one looks at the shape of $\ln (m)$ which can 
be considered a sort of ``order parameter'', near the 
critical point $a_c$ (see Fig.(2)).
Moreover, from very general arguments \cite{fln},  
we expect that the typical length
$\xi_{\perp}$ within which the polymer is confined around
the potential, diverges at the critical point as 
$\xi_{\perp}\sim |a-a_c|^{-\nu_\perp}$ with a given characteristic
exponent.  
In a sense, the variable $\nu$ appearing in (49), can be considered
a sort of external control parameter for the system 
described by (44) at equilibrium.

In order to calculate the critical exponent $\nu_\perp$ we can 
introduce the generating function $G(\lambda)$ associated to
 $P(\nu)$, as 
\be
G(\lambda)=\left\langle e^{\lambda\nu}\right\rangle
= \sum_{\nu=0}^d P(\nu)e^{\lambda\nu}.
\ee
The various momenta $\zeta_m=\langle \nu^m\rangle$
 can be calculated from $G(\lambda)$ in the usual way: 
$\zeta_m=\partial_\lambda^{(m)} G(\lambda)
|_{\lambda=0}$.
In order to study the behavior of $\xi_\perp$ we need
the knowledge  of the fluctuations of the polymer around the 
origin, and therefore we need the second cumulant 
$\mu_2=\zeta_2-\zeta_1^2$. 
We thus  calculate  the connected generating function
$\Gamma(\lambda)=\log G(\lambda)$, since 
$\mu_m=\partial_\lambda^{(m)} \Gamma(\lambda)
|_{\lambda=0}$.

From the above exact formula, we can write that
\begin{eqnarray}
G(\lambda)&=&(\eps-1)\int_0^\infty du\; 
\exp \{
-(\eps-1+dt)u +d\ln [\cosh(ut)(1+K\tanh(ut))] \} 
\nonumber \\
&=&(\eps-1)\frac{d}{\alpha}
\int_0^\infty dx\;  
\exp\left\{
d\left[ \ln [\cosh(x)(1+K\tanh(x))] 
-\frac{\eps-1+dt}{\alpha}x\right]\right\},
\end{eqnarray}
where $K=e^{\lambda}$ and $x=ut$.
If we are interested in the large $d$ behavior , 
the integral can be estimated by saddle  point  methods. 
The function at the exponent is maximum in $x=0$ if $\eps>1$,  
and then
\begin{eqnarray}
G(\lambda) &\simeq& (\eps-1)\frac{d}{\alpha} 
\int_0^\infty dx\;  
\exp\left\{d\left[ \left(K-\frac{\eps-1+dt}{\alpha}\right)
+\frac{1-K^2}{2}x^2\right]\right\} \nonumber \\
&\simeq& (\eps-1)\left[\frac{1}{\eps-1+\alpha(1-K)}
+\frac{(1-K^2)\alpha^2}
{((\eps-1+\alpha)/\alpha-K)^3} \frac{1}{d} \right].
\end{eqnarray}
Corrections to the previous formula are of the order $O(1/d^2)$.
By applying the definition of $\Gamma(\lambda)$,
 we finally  find that
\begin{eqnarray}
\mu_1 &=&\frac{dt}{\eps-1}-\frac{2(dt)^5}{(\eps-1)^2}\frac{1}{d} +
O\left(\frac{1}{d^2}\right) \nonumber \\
\mu_2&=&\frac{(dt)^2}{(\eps-1)^2}-\frac{2(dt)^5(\eps-1+4dt)}{
(\eps-1)^3}\frac{1}{d}+O\left(\frac{1}{d^2}\right).
\end{eqnarray}
As expected, the fluctuations around the average have a power-law
  divergence at the
critical point $\eps=1$. Since $\eps$ goes to 1 linearly
with $a\rightarrow a_c$,  we deduce that
the critical exponent is $\nu_\perp=1$ at $d\rightarrow \infty$.

It is also interesting to look at the shape of the partition function 
in $\nu$. From the biological point of view,   it tells us how
mutants of a given MS are distributed around it  to form a quasi-species.
If we restrict ourselves, for simplicity, to the leading term in $1/d$ 
in eq.(52),
we must inverse transform  it to get the  real space solution at first
order. 
To simplify the calculation,
 we assume that $\lambda=i\eta$ is a complex number,
and this allows us to write
\be
P^{(1)}(\nu)=(\eps-1)\int_{-\infty}^\infty d\eta e^{-i\eta\nu}
\frac{1}{\eps-1+dt(1-e^{i\eta})}.
\ee
By analytic continuation in the complex plane,  $\eta=z$ becomes 
a complex variable and the resulting integral can be calculated by means
of residues theorem. 
The integrand has a simple pole  at $z^*=-i\ln [1+(\eps-1)/(dt)]$ and 
to apply Cauchy's lemma we must close the integration path in the
semiplane $\Im\{z\}<0$.  After having calculated the residue in $z^*$,
we find that ${\rm Res}(z^*)=-i\alpha(1+(\eps-1)/\alpha)^{-\nu-1}$.
Hence,
\begin{eqnarray}
P^{(1)}(\nu)&=&{\cal N}\frac{2\pi}{dt}(\eps-1)\left(
1+\frac{\eps-1}{dt}\right)^{-(\nu+1)} \nonumber \\
&=&{\cal N}\frac{2\pi(\eps-1)}{\eps-1+dt}\exp\left[-\nu\log\left(
1+\frac{\eps-1}{dt}\right)\right], 
\end{eqnarray}
where ${\cal N}$ is a normalization factor. It can be easily calculated
by noting that  the sum involves a truncated  
geometric series:
\begin{eqnarray}
\sum_{\nu=0}^d P^{(1)}(\nu)&=&2\pi \frac{\eps-1}{\eps-1+dt}
\frac{(1-a^{d+1})}{1-a}, \nonumber \\
a^{-1}&=&\left(1+\frac{\eps-1}{dt}\right).
\end{eqnarray}
The partition function shows an exponential decay as a function of
$\nu$. The {\it mass gap} \cite{fln} is therefore given by 
$\log(1+(\eps-1)/\alpha) \simeq (\eps-1)/\alpha$, close to the
phase transition.  
Since the transversal correlation length is usually defined
as the inverse of the mass gap, we again recover the result that,  
at the critical point,  $\nu_\perp=1$.

A more refined expression of the partition function at large $d$
can be obtained by directly considering a saddle  point approximation
of   (49).
Without entering into mathematical details (similar to those
employed in previous calculations), we see that the integral
in (49) is dominated by the region close to $u^*=0$.
By expanding the  integrand around $u^*$ and integrating term 
by term,  we finally find 
\begin{eqnarray}
P(\nu)&=&(\eps-1)
\left(
\begin{array}{c} d \\ \nu
\end{array}\right)\left(\frac{\alpha}{d(\eps-1+\alpha)}\right)^\nu
\nonumber \\
&\times&\left[\frac{1}{\eps-1+\alpha}
\Gamma(\nu+1) \right.
+ \left.\frac{\alpha^2}{2d(\eps-1+\alpha)^3}
\Gamma(\nu+3)\frac{1}{d}  \right. \nonumber \\
&+& \left.O\left(\frac{1}{d^2} \right)
\right].
\end{eqnarray}
In Fig.(4) we compare this approximate result (by  only retaining 
the first term in parenthesis)  with
$P^{(1)}(\nu)$ given by (55).
The coincidence of the two curves is good up to $d\sim\nu$, i.e.
in the physical range (recall that by definition $\nu \le d$).
In fact it is possible to show  that  (57)  is 
a monotonic increasing function of $\nu$ for $\nu\gg d$, while
the exact function is always decreasing. 
The minimum of  the approximating function is found indeed
for $\nu\sim d$.
More precisely, if we only 
take the first term in (57),  a rapid inspection shows that 
it can be rewritten as
\begin{eqnarray}
P^{(1)}(\nu)&=&(\eps-1)\frac{d!}{(d-\nu)!}\frac{\alpha}{d(\eps-1+\alpha)}
\left(\frac{\alpha}{d(\eps-1+\alpha)}\right)^\nu \nonumber \\
&\sim&{\cal N}'
\frac{(\eps-1)}{\eps-1+\alpha}\left(1+\frac{\eps-1}{\alpha}
\right)^{-\nu},
\end{eqnarray}
the last approximation being valid if $\nu \ll d$. 
We then see that, apart from inessential  factors, 
eq. (55) and (58)  give the same result  only if $\nu \ll d$.
At larger $\nu$,  (58) shows
the presence of power-law corrections in the exponential decay of
the partition sum.

\section{Comparison with previous results and conclusions.}

We are now in position to compare our result with the general approach
by coming back to the usual ``quasi-species''  notation.
The copying fidelity in a  given 
reproduction process  has been defined 
in our model by $1-dt$, while in the original work \cite{eigen}
it was indicated by $q^d$ (see also eq. (7)). 
Therefore,  the first result of our work has been to show that 
the critical threshold for quasi-species formation is given by
\be
a_c=\frac{1}{1-dt}=q^{-d}, \qquad d_c=-\frac{\log a}{\log q}
\ee
which coincides with (8).

Let us now  discuss about  similarities and 
differences between   our mapping  and the 
previous approaches. 
In the above citated  work, Leuth\"ausser  introduced a mapping of
the Eigen's model to a  system at equilibrium. 
In a few words, the mapping goes as follows. Let us consider  again 
eq.(4)
with discretized time $k$, representing ``generations''of 
macromolecules.
If we define the vector ${\bf X}(k)=(x_1(k),x_2(k),\cdots,x_{2^d})$,
representing the set of the relative  concentrations of the
macromolecules at time $k$, Eigen's model can be  easily rewritten as
\be
{\bf X}(k)={\bf W}^k {\bf X}(0).
\ee
As in our case, the problem is then reduced to a linear system
associated to  ${\bf W}$. This matrix, actually,
can be though of as a transfer matrix of  an equilibrium system.
In fact, if one considers only binary sequences $I_j$, each of them
is made of  $d$ Ising spins 
$(\sigma_1,\sigma_2,\cdots,\sigma_d)$,  and the evolution of 
the system can be represented in a square lattice geometry.
One side of the lattice (made, for instance, of different
rows) has a length  equal to that  ($d$) of the sequences, while 
the other one is semi-infinite in one direction, as  each column 
can be associated to the state of the system at time $k$.
The final state, in this geometry, is therefore associated to
the  edge properties on the lattice, which  represents the state
of the system after $N$ generations. 
If each site along the binary chain is exactly copied with 
probability $q$, independently  from  other sites, 
the   replication matrix takes the form
\be
W_{ij}=A_jq^d\left(   \frac{1-q}{q}\right)^
{\left(d-\sum_{k=1}^d \sigma_k^i\sigma_k^j \right)/2}.
\ee 
This represents a transfer matrix of a two dimensional  Ising-like
system with nearest neighbors interactions along the ``time direction''.
The Hamiltonian corresponding to (61) has however a very
complicated  mathematical form
\begin{eqnarray}
-\beta{\cal H}&=&-\sum_{i=0}^{N-1} \left[\beta\sum_{j=1}^d
\sigma_j^i\sigma_j^{i+1}+\ln A(I_i)\right]\nonumber \\
&+&\frac{Nd}{2}\ln [q(1-q)],
\end{eqnarray}
that, in practice, does not allow for any analytical approach.
Tarazona \cite{t92} numerically solved the system for various
fitness landscapes $A_j$, and discussed the results in respect
to the original quasi-species model.

Apart from the intrinsic difficulty in solving problems described
by hamiltonians of the kind (62),  there is a subtle problem 
contained in this formulation. 
The actual state of the system after $N$ generations, depends 
{\it only} on the structure of the layer at the edge of the 
square lattice, that is on the spin configurations at the $N$th
column. 
Therefore, as one  may expect, the error threshold transition 
cannot be fully understood in terms of the bulk  properties on the 
square lattice, as  already pointed out 
in \cite{t92}. We  thus need the complete knowledge of the 
structure of the lattice surface, and not of the bulk, to solve
the original Eigen's model.  
With the Leuth\"ausser mapping, there is no hope to accomplish
that goal, even for the simplest possible replication landscape.
The fact that the critical properties of
the quasi-species model are associated to
surface structures is, in a sense, conserved in our mapping, as we 
have also associated the  error threshold problem to the 
statistical mechanics properties of an interface-like object. 

In conclusion,  we have analyzed the  Eigen's model in the simplest
situation characterized 
by a single-peaked  fitness. The main issues of our exact solution
can be summarized in three  main points.
First, we have proved that, in the limit of infinite sequence lengths $d$,
the error threshold  phenomenon is associated 
to  a first order  critical phase transition. 
Moreover, the 
the typical amplitude  of the quasi-species  around the 
MS diverges with 
exponent $\nu_{\perp}=1$ at criticality.
Numerical simulations \cite{t92}, seem however to indicate that 
this picture no longer holds for 
more general situations. It would be extremely 
interesting to use our mapping to investigate these other
cases, as well.
Finally, we have  proved 
that the critical selective advantage for quasi-species formation 
depends exponentially on the sequence length $d$. 

We believe that, even  in more realistic situations, in which 
the fitness landscape is characterized by rough fluctuations
from point to point, and with the help of the directed polymers 
theory, the present study can be extended. 

\section*{Acknowledgments} 
We would like to thank R. Graber and Y.-C. Zhang for
useful comments and discussions.

\begin{figure}
\caption{
The maximum eigenvalue of the transfer matrix 
${\bf  T}$ plotted vs. the selective advantage $a$
for $d=100$, $t=0.003$.
 The full line has been obtained by numerical diagonalization
of the transfer matrix.  Circles represent  the  analytical result (up
to order $O(1/d^3)$, see eq.(41)). The dashed lines are the upper
and lower  bounds
for $\eps$  obtained from the transfer matrix (see text).}
\end{figure}

\begin{figure}
\caption{
$\log[m(a)]$ is plotted  vs. the selective 
advantage  $a$ for $d=100$,  $t=.003$. 
At $a_c = (1-dt)^{-1}$ the sharp jump indicates the presence
of a depinning transition (which is well defined only  at 
$d\rightarrow\infty$, as explained in the text).
Numerical diagonalization: full line. Analytical result: circles.}
\end{figure}

\begin{figure}
\caption{
The critical dimension $d_c$ plotted  vs.  $a$ for
two distinct values of $t$. Lower curve: $t=10^{-2}$;
upper curve: $t=10^{-3}$.
Full lines represent  the function
$d_c=t^{-1}(1-1/a)$ (see text). 
Circles and squares: numerical data from the transfer matrix. }
\end{figure}

\begin{figure}
\caption{Comparison between the two real-space forms of
the solution $P(\nu)$ at the first significative order in $1/d$.
The dashed  and the full lines correspond to (55) and (58),
respectively.}
\end{figure}

\end{document}